\documentclass[12pt,a4,sans]{article}
\usepackage[latin1]{inputenc}
\usepackage{amssymb}
\usepackage{amsmath}
\usepackage{amsthm}
\usepackage{amsfonts}
\usepackage[pdftex]{graphicx}
\usepackage{geometry}
\usepackage{bbold}
\usepackage{braket}
\usepackage{enumerate}
\usepackage[usenames,dvipsnames]{xcolor}
\usepackage{setspace}
\usepackage[backgroundcolor=white,linecolor=black]{todonotes}
\usepackage[in]{fullpage}
\usepackage{hyperref}
\usepackage{array}
\usepackage{cancel}
\usepackage{wrapfig}
\usepackage[font=small,labelfont=bf]{caption}
\usepackage{anysize}

\numberwithin{equation}{section}

\marginsize{2.4cm}{2.4cm}{2cm}{2cm}

\hypersetup{
colorlinks=true,
linktoc=page,
citecolor=Blue,
linkcolor=Blue,
urlcolor=Blue} 

\makeatletter 
\renewcommand{\maketitle} 
 { \begingroup \begin{center} \large {\bf \@title}
 	\vskip 5pt \large \@author \\ \vskip 5pt \@date \end{center}
   \vskip 5pt \endgroup \setcounter{footnote}{0} }
\makeatother 

\newcommand{\comments}[1]{}
\newcommand{\la}{\langle}
\newcommand{\ra}{\rangle}
\newcommand{\tl}{\widetilde\lambda}

\newcommand{\A}{\mathcal{A}}

\newcommand{\N}{\mathcal{N}}

\newcommand{\Tr}{\text{Tr}}

\renewcommand{\b}[1]{\braket{#1}}

\renewcommand{\O}{\mathcal{O}}

\newcommand{\be}{\begin{equation}}
\newcommand{\ee}{\end{equation}}

\def\beqa{\begin{eqnarray}}
\def\eeqa{\end{eqnarray}}
\def\beq{\begin{equation}}
\def\eeq{\end{equation}}

\def\Tr{{\rm Tr}}
\def\one{\mbox{1 \kern-.59em {\rm l}}}

\def\cA{{\cal A}} \def\cB{{\cal B}} \def\cC{{\cal C}}

 \def\cN{{\cal N}} \def\cO{{\cal O}}

\def\D{\Delta}

\def\uno{\mbox{1 \kern-.59em {\rm l}}}

\def\lan{\langle}
\def\ran{\rangle}

\def\one{1\!\!1\,\,}

\def\bcomment#1{}

\def\eps{\epsilon}

\usepackage{slashed}
\usepackage{caption}

\usepackage[nottoc,notlot,notlof]{tocbibind}
\usepackage[nosort]{cite}
\usepackage{color}
\usepackage{bbm}
\usepackage{parskip}
\usepackage{mathtools}

\graphicspath{{./Images/}}

\long\def\symbolfootnote[#1]#2{\begingroup%
\def\thefootnote{\fnsymbol{footnote}}\footnote[#1]{#2}\endgroup}

\begin{document}

\begin{flushright}
QMUL-PH-15-04\\
HU-EP-15/09
\end{flushright}

\vspace{20pt}

\begin{center}

{\Large \bf Integrability and Unitarity   }\\

%
\vspace{45pt}

{\mbox {\bf  Andreas Brandhuber$^{a}$,  Brenda Penante$^{a}$, Gabriele Travaglini$^{a,b}$ and  Donovan Young$^{a}$}}%
\symbolfootnote[4]{
{\tt  \{ \tt \!\!\!a.brandhuber, b.penante, g.travaglini, d.young\}@qmul.ac.uk}
}

\vspace{0.5cm}

\begin{quote}
{\small \em
\begin{itemize}
\item[\ \ \ \ \ \ $^a$]
\begin{flushleft}
Centre for Research in String Theory\\
School of Physics and Astronomy\\
Queen Mary University of London\\
Mile End Road, London E1 4NS, United Kingdom
\end{flushleft}
\item[\ \ \ \ \ \ $^b$]
Institut f\"{u}r Physik und IRIS Adlershof\\
Humboldt-Universit\"{a}t zu Berlin\\
Zum Gro{\ss}en Windkanal 6, 12489 Berlin, Germany

\end{itemize}
}
\end{quote}


\vspace{40pt}

{\bf Abstract}
\end{center}

\vspace{0.3cm} 

\noindent

\noindent
We show how generalised unitarity can be used to determine the
one-loop dilatation operator in $\cN\!=\!4$ super Yang-Mills. Our
analysis focuses on two sectors, namely the bosonic $SO(6)$ sector and
the $SU(2|3)$ sector. The calculation is performed on shell,
with no off-shell information introduced at any stage. In this way, we
establish a direct connection between scattering amplitudes and the
dilatation operator of the $\cN\!=\!4$ theory.

\setcounter{page}{0}
\thispagestyle{empty}
\newpage



%


\section{Introduction}

In this paper we continue the study initiated in
\cite{Brandhuber:2014pta}, whose ultimate goal is to relate scattering
amplitudes in $\cN=4$ super Yang-Mills (SYM) theory, and their
properties, to the dilatation operator in the same theory. In a sense,
we retrace the history of the developments in the calculation
of loop amplitudes triggered by Witten's twistor string theory
\cite{Witten:2003nn}: In \cite{Brandhuber:2014pta} we employed MHV
diagrams \cite{Cachazo:2004kj} at loop level \cite{Brandhuber:2004yw}
in order to compute the dilatation operator at one loop; here, we
proceed to apply a powerful variant of unitarity
\cite{Bern:1994zx,Bern:1994cg} known as generalised unitarity
\cite{Bern:1997sc,Britto:2004nc} which, as we shall see, allows for an
even more efficient calculation of the dilatation operator.  As
generalised unitarity turned out be more practical than loop MHV
diagrams, we will see how our use of generalised unitarity will
further simplify the already remarkably simple calculation of the
dilatation operator performed with MHV rules.

 The use of unitarity in deriving the dilatation operator is welcome
 also from a conceptual point of view, since the only ingredients of
 the calculation are on-shell amplitudes -- with no off-shell
 information being introduced. This supports the hope that using this
 approach one may be able to connect directly the amplitudes and their
 hidden structures and symmetries to the integrability of the
 dilatation operator in $\cN=4$ SYM.

At one loop, the no-triangle property \cite{Bern:1994zx} of the
one-loop S-matrix of $\cN=4$ SYM implies that maximal cuts employed in
\cite{Britto:2004nc} are enough to completely determine all amplitudes
of the theory. Similarly, we identify certain quadruple cuts which are
sufficient to determine the dilatation operator at one loop.  The
reason why this is correct lies in the simplicity of the object under
consideration, namely a two-point function, or a single-scale object
in momentum space. At one loop there are precisely four fields to
be connected, which explains why quadruple cuts are enough.

In more detail, we will focus on the dilatation operator in the
$SO(6)$ and $SU(2|3)$ sectors of $\N=4$ SYM, which we will derive by
computing the two-point functions $\la \O(x_1)\bar{\O}(x_2)\ra$ of the
appropriate composite operators. The $SO(6)$ sector was studied in
\cite{Minahan:2002ve}, where the connection to integrable spin chains
was first made, and is closed (only) at one loop. The $SU(2|3)$ sector is
closed to all loops and was considered first in \cite{Beisert:2003ys}.
At one loop and in the planar limit, only contractions of pairs of
adjacent fields survive. In all cases we consider, the result of our
calculation turns out to be proportional to the quadruple cut of the
same integral appearing in the original one-loop calculation performed
by Minahan and Zarembo. This integral is given by
\begin{align}
\label{eq:X-position}
I(x_{12})\,=\,\int\!\!d^Dz \ \D^2 (x_1 - z) \, \D^2(x_2-z)\, , 
\end{align}
where $x_{12}:= x_1 - x_2$ and  
\beq
\D  (x)  \ := \ 
- {\pi^{2 - {D\over 2}} \over 4 \pi^2}
\Gamma \Big({D\over 2} - 1 \Big)
 {1\over (-x^2+ i \varepsilon)^{{D\over 2} - 1}}
\ , 
\eeq
is the scalar propagator in $D$ dimensions. In momentum space, it
appears as the Fourier transform of the simplest single-scale
integral, namely a double bubble,
\begin{align}
\label{eq:X-momentum}
\begin{split}
I(x_{12})\,=&\,\int\!\prod_{i=1}^4 \frac{d^D
  L_i}{(2\pi)^D}\frac{e^{i(L_1+L_2)\cdot x_{12}}}{L_1^2\,L_2^2\,L_3^2\,L_4^2}\, (2 \pi)^D \, \delta^{(D)}\Big(\sum_{i=1}^4L_i\Big)\ \\
=&\,\int\!\frac{d^D L}{(2\pi)^D}\; e^{iL\cdot x_{12}}\int\!\frac{d^D L_1}{(2\pi)^D}\frac{d^D L_3}{(2\pi)^D}\frac{1}{L_1^2\,(L-L_1)^2\,L_3^2\,(L+L_3)^2}\ , 
\end{split}
\end{align}
\begin{figure}[h]
\centering
\includegraphics[width=0.3\linewidth]{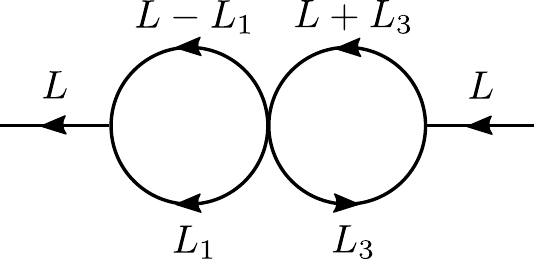}
\caption{\it The double-bubble integral relevant for the computation of $I(x_{12})$.}
\label{fig:double-bubble}
\end{figure}
\\ where $L:= L_1 + L_2$. By using the quadruple cut where the
propagators with momenta $L_1$, $L_2 := L-L_1$, $L_3$ and $L_4 :=
L+L_3$ are put on shell, we will be able to identify the coefficient
of this double bubble in all relevant cases -- without ever performing
an integral. The cut double bubble can then be lifted to a full
integral, and by picking its ultraviolet (UV) divergence
$I(x_{12}) \bigr|_{\rm UV}$, 
\beq
\label{l-uv-d}
 I(x_{12}) \bigr|_{\rm UV} \ = \ {1\over \epsilon}\cdot {1\over
  8 \pi^2} \cdot {1\over (4 \pi^2 x_{12}^2)^2} \ , 
  \eeq 
  we can
immediately write down the dilatation operator.  Let us
also mention that other applications of unitarity to the calculation
of $n$-point correlators and correlation functions of Wilson lines have appeared in
\cite{Engelund:2012re, Laenen:2014jga, Engelund:2015cfa}. It is worth stressing the
two key reasons why generalised unitarity is particularly powerful for
the case considered here, namely that of the two-point function. First, 
as we have already mentioned, quadruple cuts are precisely the right set
of cuts to identify the relevant loop integral; furthermore, the
relevant integrals are guaranteed to have a single scale. The cut
integral can then be lifted to a full loop integral without
introducing spurious discontinuities, in complete analogy to the case
of splitting amplitudes studied in \cite{Kosower:1999xi}.
Finally, we also note that our approach to the computation of the dilatation operator differs from that of 
\cite{Wilhelm:2014qua, Nandan:2014oga}  in that no infrared divergences appear at any stage in 
our calculation.

The rest of the paper is organised as follows. In the next section we
use generalised unitarity to obtain the dilatation operator in
the  $SO(6)$ sector. In Section 3 we move on to the
$SU(2|3)$ sector. This case is particularly interesting as it involves
fermions as well as scalars. There are several contributions to
consider and the structure of the dilatation operator is more
elaborate than in the pure scalar sector, hence our tests are more
stringent.  Finally in Section 4 we compare recent on-shell and twistorial approaches for the calculation of the dilatation operator, and also make a few suggestions for future work.

\section{The dilatation operator in the $SO(6)$ sector}

In this section we will compute the dilatation operator  of $\cN=4$ SYM in  the $SO(6)$ sector using generalised unitarity. This calculation was recently performed in \cite{Koster:2014fva} and \cite{Brandhuber:2014pta} using MHV diagrams in twistor space and momentum space, respectively. Here we depart from these off-shell approaches in favour of a fully on-shell calculation.

\begin{figure}[ht]
\begin{center}
\scalebox{0.9}{\includegraphics{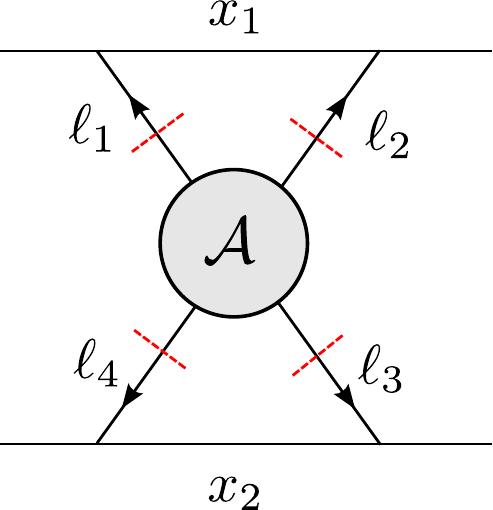}}
\end{center}
\caption{\it
The single cut diagram contributing to the dilatation operator at one loop. 
}
\label{cut-diagram}
\end{figure}

Operators in the $SO(6)$ sector have the form 
\beq
\cO_{A_1 B_1, A_2 B_2, \ldots, A_L B_L} (x) \ := \ \Tr \big( \phi_{A_1 B_1} (x) \cdots  \phi_{A_L B_L} (x) \big) \ . 
\eeq
At one loop and in the planar limit, it is sufficient to consider contractions of pairs of adjacent fields 
(in colour space). The relevant part of each  operator is then 
\begin{align}
\label{eq:operators}
\begin{split}
\O(x_1)\,&=\,\cdots \phi^a_{AB}(x_1)\phi^b_{CD}(x_1)\cdots (T^a T^b)^i_{\phantom{i}j}\ ,\\
\bar{\O}(x_2)\,&=\,\cdots \phi^c_{A'B'}(x_2)\phi^d_{C'D'}(x_2)\cdots (T^c T^d)^{l}_{\phantom{i'}m}\ .
\end{split}
\end{align}
The calculation is then effectively equivalent to that of the following two-point function  
\begin{align}
\left\langle (\phi^a_{AB} \phi^b_{CD})(x_1)(\phi^c_{A^\prime B^\prime} \phi^d_{C^\prime D^\prime})(x_2)  \right\rangle \ ,  
\end{align}
whose expected  structure  is
\beq
\label{ABC}
\big\langle (\phi_{AB} \phi_{CD})(x_1)(\phi_{A^\prime B^\prime} \phi_{C^\prime D^\prime})(x_2) \big\rangle =  \A  \epsilon_{ABCD} \epsilon_{A^\prime B^\prime C^\prime D^\prime}  +  \mathcal{B}  \epsilon_{AB A^\prime B^\prime} \epsilon_{CD  C^\prime D^\prime}  + \mathcal{C}   \epsilon_{AB C^\prime D^\prime} \epsilon_{A^\prime B^\prime CD} 
\ . 
\eeq
These three terms are usually referred to  as  trace, permutation and identity. The dilatation operator can then be read off from the UV divergences of  \eqref{ABC}, hence  we only need to compute  the UV-divergent parts 
$\mathcal{A}_{\rm UV}$,  $\mathcal{B}_{\rm UV}$, $\mathcal{C}_{\rm UV}$ of   the coefficients 
$\mathcal{A}$,  $\mathcal{B}$ and  $\mathcal{C}$.
These are expected to be equal to \cite{Minahan:2002ve} 
\beq
\A_{\rm UV} \ = \ {1\over 2}\, , \qquad  \mathcal{B}_{\rm UV} \ = \ -1 \, , \qquad \mathcal{C}_{\rm UV} \ = \ 1 \ . 
\eeq
As in \cite{Brandhuber:2014pta}, we  choose the following $SU(4)$ assignments in \eqref{eq:operators} as representatives of these three  flavour structures:
\begin{align}
\label{table1}
\begin{array}{c|cc}
 & ABCD & A'B'C'D' \\
 \hline
 \text{Tr} &1234 & 2413 \\
 \mathbb{P} & 1213 & 3424\\
 \uno & 1213 & 2434
\end{array}
\end{align}
For each case there is a single cut diagram to consider. The integrand
is constructed with four cut scalar propagators with momenta $L_i$,
$i=1, \ldots , 4$,  and one on-shell amplitude, as shown in Figure \ref{cut-diagram}. The operators are
connected to the amplitude via appropriate form factors, which in the
scalar case are simply
\begin{align}
\label{eq:form-factor-scalars}
\begin{split}
F_{\phi^a \tilde\phi^b}(\ell_1^{\phi^{a'}},\ell_2^{\tilde\phi^{b'}};L)\,&:=\,
\int\!\!d^4x \  e^{iL\cdot x}\ 
 \big\langle \, 0\, |(\phi^a \tilde\phi^b)(x)|\, \phi^{a'}(\ell_1),\tilde\phi^{b'}(\ell_2)\, \big\rangle\\
\,&=\, (2 \pi)^4 \delta^{(4)}\big(L-\ell_1-\ell_2\big)\,  \delta^{aa'}\delta^{bb'}\  ,
\end{split}
\end{align}
where we have used $\phi$ and $\tilde\phi$ to denote two scalar fields
having distinct $R$-symmetry indices as is sufficient for our
purposes, see (\ref{table1}). Note that the $\ell_i$ represent the
on-shell (cut) versions of the loop momenta $L_i$.

The relevant amplitudes for the three flavour assignments considered in  \eqref{table1} are:%
\footnote{In the following expressions we omit a factor of $g_{\rm
    YM}^2$, which will be reintroduced at the end of the calculation.}
\beqa
{\rm Tr}: \qquad 
A(1^{\phi_{12}},4^{\phi_{13}},3^{\phi_{24}},2^{\phi_{34}}) &=&   \dfrac{ \lan 13\ran \lan 24\ran}{\lan 12 \ran \lan 34 \ran} \, , 
 \\ \cr
\mathbb{P}: \qquad 
A(1^{\phi_{12}},4^{\phi_{24}},3^{\phi_{34}},2^{\phi_{13}}) &=&  -1 \, , 
 \\ \cr
\uno: \qquad 
A(1^{\phi_{12}},4^{\phi_{34}},3^{\phi_{24}},2^{\phi_{13}}) &=&    \dfrac{ \lan 13\ran \lan 24\ran}{\lan 23 \ran \lan 14 \ran} \, .
\label{sss}
\eeqa 
Three observations are in order here. First, we note that the
same integrands as in the approach of \cite{Brandhuber:2014pta} have
appeared, with the important difference that, in that paper, the
spinors associated with the on-shell momenta are given by the
appropriate off-shell continuation for MHV diagrams. Here the spinors
for the cut loop momenta do not need any off-shell continuation.
Furthermore, for the case of the $\mathbb{P}$ integrand there is
obviously no difference between the two approaches, and the resulting
integral is given by a double bubble where all the four propagators
are cut.  In the other two cases, this integral is dressed by the
appropriate amplitude.  Finally, we note that the colour factor associated with all diagrams is 
obtained from the contraction 
\beq
\label{NN}
\cdots (T^{b}T^a)^{i}_{\, j} \cdots  \Tr (T^a T^b T^c T^d) \cdots (T^d T^c)^l_{\, m} \cdots = \cdots N^2 \delta^i_{m} \delta ^l_j \cdots\ , 
\eeq
 where the trace arises from the amplitude and the  factors $\cdots (T^{b}T^a)^{i}_{\, j} \cdots $ and $\cdots (T^d T^c)^l_{\, m} \cdots$ from the operators (and we indicate only generators corresponding to the fields being contracted). 
We now proceed to construct the relevant integrands.

\subsection*{The trace integrand} 

In this case the relevant amplitude (which multiplies four cut
propagators) can be rewritten as\footnote{We define $\Tr_+(abcd)
  := \b{ab}[bc\,]\b{cd}[da]$. }
\begin{align}
\frac{\b{13}\b{24}}{\b{12}\b{34}}\,=\,\frac{\Tr_+(\ell_1\,\ell_3\,\ell_4\,\ell_2)}{(\ell_1+\ell_2)^2(\ell_3+\ell_4)^2}
\,=\,-\frac{2(\ell_1\cdot\ell_3)}{L^2}
\ , 
\end{align}
where we have used  $\ell_1+\ell_2=-(\ell_3+\ell_4):= L$. Having rewritten the amplitude in terms of products of momenta, we lift the four cut momenta off shell. The resulting integral has  the  structure of a product of two linear bubbles,  
\begin{align}
\label{eq:trace}
-\frac{2}{L^2}\int\!\frac{d^D L_1}{(2\pi)^D}\frac{L_1^\mu}{L_1^2\,(L-L_1)^2}\int\! \frac{d^D L_3}{(2\pi)^D}\frac{L_{3\,\mu}}{L_3^2\,(L+L_3)^2}\ .
\end{align}
Using the fact that 
\beq
\int\!{d^D K \over (2 \pi)^D} {K^\mu \over K^2 (K \pm L)^2} \ = \ \mp {L^\mu\over 2} {\rm Bub}(L^2)  \, ,
\eeq
where (in Euclidean signature) 
\beq
{\rm Bub} (L^2) := \int\!{d^D K \over (2 \pi)^D} {1 \over K^2 (K + L)^2}\, = \, {1\over (4 \pi)^{D\over 2}} { 
\Gamma ( 2 - {D \over 2} ) \Gamma^2 ({D\over 2} - 1 ) \over \Gamma (D - 2)} (L^2)^{{D\over 2} - 2} 
\ , 
\eeq
we find that  \eqref{eq:trace} is equal to  $1/2$ times a double bubble. 
Using \eqref{l-uv-d} 
 we finally get  $\A_{\rm UV}=1/2$. Note that in arriving at this result we have performed a Fourier transform to position space using 
 \beq
 \int\!{d^D p\over (2\pi)^D} \, {e^{i p \cdot x} \over (p^2)^s} \ = \   { \Gamma( {D\over 2} - s) \over 4^s \, \pi^{D\over 2}\, \Gamma (s) } \, {1\over (x^2)^{ {D\over 2}  - s}}
 \ . 
 \eeq
Furthermore, in  the definitions of  $\cA_{\rm UV}$, $\cB_{\rm UV}$, and $\cC_{\rm UV}$ (and  in general of any other UV-divergent coefficient in the rest of the paper),  a factor of $\lambda/(8 \pi^2) \times \big(1 / ( 4 \pi^2 x_{12}^2)\big)^2 \times (1/ \epsilon)$ will always be understood,  with $\lambda := g_{\rm YM}^2 N$.

\subsection*{The $\mathbb{P}$ integrand}
No calculation is needed in this case, and the result is simply given by minus a cut double-bubble integral. Lifting the cut integral to a full loop integral  we get  $\mathcal{B}_{\rm UV}=-1$.

\subsection*{The $\uno$ integrand}

\begin{figure}[h]
\centering
\includegraphics[width=0.31\linewidth]{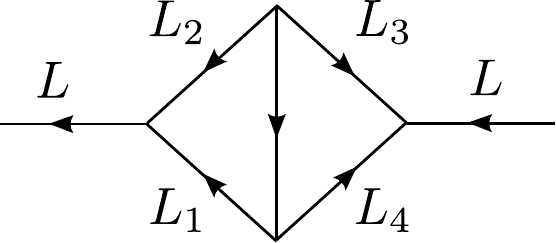}
\caption{\it The finite integral corresponding to the term in \eqref{sem}. This integral is irrelevant for the calculation of the dilatation operator.}
\label{finiteintegral}
\end{figure}

The relevant amplitude in this case is
\begin{align}
\label{eq:amp-uno}
\frac{\b{13}\b{24}}{\b{23}\b{14}}\,=\,1+\frac{\b{12}\b{34}}{\b{23}\b{14}}\ .
\end{align}
Thus the first term in \eqref{eq:amp-uno} gives the cut double-bubble integral, whereas we can use  on-shell identities to rewrite the second term as
\begin{align}
\label{sem}
\frac{\b{12}\b{34}}{\b{23}\b{14}}\,=\,\frac{\b{12}\b{34}[34]}{\b{23}\b{14}[34]}=-\frac{L^2}{2(\ell_1\cdot\ell_4)}\ . 
\end{align}
Lifting the cut propagators of the second integral to full propagators,  it is immediate to see  that this term produces the integral represented in Figure \ref{finiteintegral}. This integral is finite in four dimensions and  
thus does not contribute to $\mathcal{C}_{\rm UV}$. We then conclude that $\mathcal{C}_{\rm UV}=1$. 
\\
A comment is in order here. In principle an ambiguity is still present corresponding to an integral such as that of Figure \ref{finiteintegral} but with one of the four propagators $L_1, \ldots , L_4$ collapsed (say $L_4$), which is UV divergent. This integral can be excluded by looking at a triple cut corresponding to  cutting the  propagators $L_1$, $L_3$ as well as the middle propagator  in  Figure~\ref{finiteintegral}.

For later convenience, we explicitly write down the form of the UV-divergent part of the correlator \eqref{ABC}, 
\begin{align}
\begin{split}
& \left.\big\langle \big(\phi_{AB} \phi_{CD}\big)^i_{\, j} (x_1)\big(\phi_{A^\prime B^\prime} \phi_{C^\prime D^\prime}\big)^l_{\, m}(x_2)  \big\rangle\right|_{\rm UV} \\
=\, & {1\over \eps} \cdot {\lambda \over 8 \pi^2} 
\Big( \Delta^2 (x_{12})
  \delta^i_m \delta ^l_j\Big)  \, \Big( {1\over 2}  \,\epsilon_{ABCD} \epsilon_{A^\prime B^\prime C^\prime D^\prime}  -  \epsilon_{AB A^\prime B^\prime} \epsilon_{CD  C^\prime D^\prime}  +  \epsilon_{AB C^\prime D^\prime} \epsilon_{A^\prime B^\prime CD} \Big)
\ . 
\end{split}
\end{align}
In terms of a spin-chain Hamiltonian, this can be represented as \cite{Minahan:2002ve}
\beq
\label{pss}
H_2 = {\lambda \over 8 \pi^2} \Big( {1\over 2} \, {\rm Tr} \, +\, \uno \, -\, \mathbb{P} \Big)
\ . 
\eeq

\section{The dilatation operator in the $SU(2|3)$ sector}

In this section we consider the closed $SU(2|3)$ sector. This is particularly interesting, as it involves also fermions. Indeed,  operators in this sector are formed with letters taken from the set  
 $\big\{\psi_{1\,\alpha}, \phi_{1A}\big\}$, with $\alpha=1,2$ and $A=2,3,4$. We thus have one  fermion and three scalar fields. 
The dilatation operator in this sector was derived in \cite{Beisert:2003ys}. 
Its expression is given by
\beq
\label{dilsu23}
\hspace{-0.1cm} H_2=\dfrac{\lambda}{8\pi^2}\bigg[
  \left\{\begin{smallmatrix}A\,B\\\cr A\,B\end{smallmatrix}\right\} \,
  -\, \left\{\begin{smallmatrix}A\,B\\\cr
  B\,A\end{smallmatrix}\right\} \, + \,
  \left\{\begin{smallmatrix}A\,\beta\\\cr
  A\,\beta\end{smallmatrix}\right\}\,+\,
  \left\{\begin{smallmatrix}\alpha\,B\\\cr
  \alpha\,B \end{smallmatrix}\right\} \, - \, \Big(
  \left\{\begin{smallmatrix}A\,\beta\\ \cr
  \beta\,A\end{smallmatrix}\right\}\,+\,
  \left\{\begin{smallmatrix}\alpha\,B\\ \cr
  B\,\alpha \end{smallmatrix}\right\}\Big) \,+\,
  \left\{\begin{smallmatrix}\alpha\,\beta\\ \cr
  \alpha\,\beta\end{smallmatrix}\right\}
  \,+\,\left\{\begin{smallmatrix}\alpha\,\beta\\ \cr
  \beta\,\alpha\end{smallmatrix}\right\} \bigg] , \eeq where in this
notation the three scalar fields are labelled by $A, B=2,3,4$. In the
following we are going to rederive \eqref{dilsu23} using an
application of generalised unitarity.

As for the $SO(6)$ case, in the planar limit only contractions between
nearest-neighbour fields in $\O(x_1)$ and $\bar{\O}(x_2)$ have to be
considered. The first two terms on the right-hand side of 
\eqref{dilsu23} denote the scalar identity $\uno$ and permutation
$\mathbb{P}$ structures already familiar from the $SO(6)$ case (the
trace structure is absent given the restricted choice of scalar
letters).  The novelty is that now we have to consider two additional
types of contractions: scalar-fermion $\to$ scalar-fermion, and
two-fermion $\to$ two-fermion, as indicated in the remaining terms in
\eqref{dilsu23}.

\subsection*{Scalar-fermion $\to$ scalar-fermion}

In this case we are interested in a fermion field $\psi_{1\,\alpha}$ and one of the scalars $\phi_{12},\,\phi_{13},$ or $\phi_{14}$. Without loss of generality we will consider $\phi_{12}$. There are two cases to consider, 
\beq   
\mathbf{U}:
\qquad \big\la( 
\phi^a_{12} \psi^b_{1\,\alpha})(x_1)(\psi^c_{234\,\dot{\alpha}}\phi^d_{34})(x_2)\big\ra\, , 
\eeq
and 
\beq
\mathbf{S}:
\qquad
\big\la( \phi^a_{12} \psi^b_{1\,\alpha})(x_1)(\phi^c_{34}\psi^d_{234\,\dot{\alpha}})(x_2)\big\ra\ , 
\eeq
where the letters {\bf U} and {\bf S} indicate whether the contractions between the two fields  are unswapped or swapped.
The relevant form factor  is 
\beq
\label{eq:form-factor-fermion-scalar}
\begin{split}
F_{ \phi^a_{12}  \psi_{1\,\alpha}^b}(\ell_1^{\phi_{12}^{a'}}, \ell_2^{\psi_{1\,\alpha}^{b'}};L)\,
&:=\,
\int\!\!d^4x \ e^{iL\cdot x} \, \la 0|( \phi^a_{12}  \psi_{1\,\alpha}^b)(x)|\phi_{12}^{a'}(\ell_1),\psi_{1}^{b'}(\ell_2)\ra\\
\,&=\, (2 \pi)^4 \delta^{(4)}\big(L-\ell_1-\ell_2\big)\lambda^2_\alpha\,\delta^{aa'}\delta^{bb'}\, ,
\end{split}
\eeq
and similarly for $\bar{\O}(x_2)$. 

We begin by considering the {\bf U} case. By 
contracting the two form factors with the four planar permutations of the full amplitude, we obtain\footnote{Two out of the six possible contractions obviously do not contribute at large $N$.} 
\begin{align}
\label{eq:contraction-fermion-scalar-1}
\begin{split}
&\lambda^2_\alpha\tl^3_{\dot{\alpha}}\,\delta^{aa'}\delta^{bb'}\delta^{cc'}\delta^{dd'}\\
\times \Big[&A( 1^{\phi_{12}}, 2^{\psi_1},3^{\psi_{234}},4^{\phi_{34}})\,\Tr(T^{a'}T^{b'}T^{c'}T^{d'}) +
 A(1^{\phi_{12}}, 2^{\psi_1}, 4^{\phi_{34}},3^{\psi_{234}})\,\Tr(T^{a'}T^{b'}T^{d'}T^{c'})\\
 - &A(1^{\phi_{12}}, 3^{\psi_{234}}, 4^{\phi_{34}},2^{\psi_1})\,\Tr(T^{a'}T^{c'}T^{d'}T^{b'}) - 
 A(1^{\phi_{12}}, 4^{\phi_{34}},3^{\psi_{234}},2^{\psi_1})\,\Tr(T^{a'}T^{d'}T^{c'}T^{b'})
 \Big]\ .
\end{split}
\end{align}
At large $N$ there is only one leading contribution, corresponding to the term with the  amplitude 
\beq
A(1^{\phi_{12}}, 4^{\phi_{34}}, 3^{\psi_{234}}, 2^{\psi_1}) \,=\,  \frac{\b{13}\b{34}}{\b{14}\b{23}}\ . 
\eeq
It is given by
\begin{align}
\begin{split}
-A(1^{\phi_{12}}, 4^{\phi_{34}}, 3^{\psi_{234}}, 2^{\psi_1}) \,\lambda^2_\beta  \tl^3_{\dot\beta}\, =  \, - \frac{(\ell_2\,\bar{\ell_1}\,\ell_3)_{\beta\dot{\beta}}}{2(\ell_1\cdot\ell_4)}\, :=   \,N_{\beta\dot{\beta}} .
\end{split}
\end{align}
The cut  integral to consider is thus 
\beq
\label{888}
I_{\beta\dot{\beta}} := \int\!\! d^4\ell_1  d^4\ell_3 \,
\delta^{(+)} (\ell_1^2) \,  
\delta^{(+)} (\ell_3^2) \,  
\delta^{(+)}\left( (L-\ell_1)^2\right) \,  
\delta^{(+)} \left((L+\ell_3)^2\right) \,  
\ \cdot \ N_{\beta\dot{\beta}}\ ,  
\eeq
where by Lorentz invariance $I_{\beta\dot{\beta}}$ must have the form 
\begin{align}
I_{\beta\dot{\beta}}\,=\,A\,L_{\beta\dot{\beta}}\ .
\end{align}
A simple PV reduction shows that the UV-divergent part of the coefficient $A$ is  equal to%
\footnote{We recall that we omit a factor of $\lambda/(8 \pi^2) \times \big(1 / ( 4 \pi^2 x_{12}^2)\big)^2 \times (1/ \epsilon)$ after Fourier transforming to position space.} 
$A_{\rm UV} =1/2$.

For the {\bf S} case, 
we get the single leading contribution to be
\begin{align}
\begin{split}
- A(1^{\phi_{12}}, 4^{\psi_{234}}, 3^{\phi_{34}}, 2^{\psi_1})\,\lambda^2_\beta  \tl^4_{\dot\beta} \,&=\,  - \frac{(\ell_2\,\bar{\ell_1}\,\ell_4)_{\beta\dot{\beta}}}{2(\ell_2\cdot\ell_3)}\, := \,\tilde{N}_{\beta\dot{\beta}} \ .
\end{split}
\end{align}
The relevant integral is now  
\beqa
\tilde{I}_{\beta\dot{\beta}}&:=&
\label{999}
\int\!\! d^4\ell_1  d^4\ell_3 \,
\delta^{(+)} (\ell_1^2) \,  
\delta^{(+)} (\ell_3^2) \,  
\delta^{(+)} \left((L-\ell_1)^2\right) \,  
\delta^{(+)} \left((L+\ell_3)^2\right) \,  
\ \cdot \ \tilde{N}_{\beta\dot{\beta}}
\nonumber \\ 
&=&
\tilde{A}\,L_{\beta\dot{\beta}}
\ , 
\eeqa
where a PV reduction shows that   $\tilde{A}=-1/2$. Note that in arriving at this result we have discarded finite integrals, which do not contribute to the anomalous dimensions (more precisely, in all calculations the only other finite integral appearing is the kite, depicted in Figure \ref{finiteintegral}).

Summarising, the scalar-fermion $\to$ scalar-fermion case  gives $\pm 1/2 \, L_{\beta\dot{\beta}}$ times a double-bubble integral, for the {\bf U}/{\bf S} case, respectively. 
This has  to be compared to the tree-level expression
\begin{align}
\label{eq:fermion-scalar-tree}
\begin{split}
I^{\rm tree}_{\beta\dot{\beta}}\,&:=\, 
\int\frac{d^DL_1}{(2\pi)^D}\frac{L_{1\,\beta\dot{\beta}}}{L^2_1(L-L_1)^2} \ = \ {1\over 2} \, L_{\beta\dot{\beta}}\, 
{\rm Bub} (L^2)  \ .
\end{split}
\end{align}
Thus for the two-scalar two-fermion case we get: 
\beq
\uno: 1 \, , \qquad 
\mathbb{P}: -1 \ ,
\eeq
and the corresponding contribution to the spin-chain Hamiltonian is%
\footnote{Here we also reinstate powers of $g_{\rm YM}^2$ from the tree-level amplitudes, of $N$, arising from colour contractions, and a factor of $1/(8\pi^2)$ arising from the UV singularity \eqref{l-uv-d} of the double-bubble integral \eqref{eq:X-momentum}.}
\beq
\label{1f1s}
{\lambda \over 8 \pi^2}\left(  \left\{\begin{smallmatrix}A\,\beta\\\cr A\,\beta\end{smallmatrix}\right\} \, - \,
 \left\{\begin{smallmatrix}A\,\beta\\\cr \beta\,  A\end{smallmatrix}\right\} \right)
 \ , 
\eeq
in agreement with the corresponding terms in  \eqref{dilsu23}.

\subsection*{Two-fermion $\to$ two-fermion}

In this case we consider the four-point correlator 
\begin{align}
\label{eq:ops-fermion-fermion}
\big\la (\psi_{1\,\alpha}^a \psi^b_{1\,\beta})(x_1)
(\psi^c_{234\,\dot{\alpha}}\psi^d_{234\, \dot\beta})(x_2)\big\ra  \ . 
\end{align}
The form factors of $\O(x_1)$ are given by
\begin{align}
\label{eq:form-factor-fermions}
\begin{split}
F_{\psi_{1\,\alpha}^a \psi^b_{1\,\beta}}(\ell_1^{\psi_{1\,\alpha}^{a'}},\ell_2^{\psi_{1\,\beta}^{b'}};L)\,&:=\,
\int\!\!d^4x \  e^{iL\cdot x}\ 
 \big\langle \, 0\, |(\psi_{1\,\alpha}^a \psi^b_{1\,\beta})(x_1)|\, \psi_{1}^{a'}(\ell_1),\psi_{1}^{b'}(\ell_2)\, \big\rangle\\
\,&=\, (2 \pi)^4 \delta^{(4)}\big(L-\ell_1-\ell_2\big)\,\cdot {1\over 2}  \big(\lambda^1_\alpha\lambda^2_\beta\, \delta^{aa'}\delta^{bb'}\, -\,  \lambda^1_\beta\lambda^2_\alpha\, \delta^{ab'}\delta^{ba'}\big)\, ,
\end{split}
\end{align}
and similarly for the form factor of $\bar\O(x_2)$. Note the factor of $1/2$ appearing because of the presence of two identical particles in the state. 
Contracting the two form factors with the four planar permutations of the full amplitude,
we get
\begin{align}
\label{eq:contraction-fermions}
\begin{split}
-\frac{1}{4}\big(&\lambda^1_\alpha\lambda^2_\beta\delta^{aa'}\delta^{bb'}- \lambda^1_\beta\lambda^2_\alpha\delta^{ab'}\delta^{ba'}\big)\big(\tl^3_{\dot\alpha}\tl^4_{\dot\beta}\delta^{cc'}\delta^{dd'}- \tl^3_{\dot\beta}\tl^4_{\dot\alpha}\delta^{cd'}\delta^{dc'}\big)\\
\times \Big[&A(1^{\psi_1},2^{\psi_1},3^{\psi_{234}},4^{\psi_{234}})\,\Tr(T^{a'}T^{b'}T^{c'}T^{d'}) - A(1^{\psi_1},2^{\psi_1},4^{\psi_{234}},3^{\psi_{234}})\,\Tr(T^{a'}T^{b'}T^{d'}T^{c'})\\
 + &A(1^{\psi_1},3^{\psi_{234}},4^{\psi_{234}},2^{\psi_{1}})\,\Tr(T^{a'}T^{c'}T^{d'}T^{b'}) - A(1^{\psi_1},4^{\psi_{234}},3^{\psi_{234}},2^{\psi_{1}})\,\Tr(T^{a'}T^{d'}T^{c'}T^{b'})
 \Big]\ .
\end{split}
\end{align}
In the large-$N$ limit we only need to keep the following terms out of those in \eqref{eq:contraction-fermions}:
\begin{align}
\label{eq:contraction-fermions-2}
\begin{split}
-{1\over 4} \bigg[ A(1^{\psi_1},2^{\psi_1},3^{\psi_{234}},4^{\psi_{234}})\,\lambda^1_\beta\lambda^2_\alpha \tl^3_{\dot\beta}\tl^4_{\dot\alpha}+ A(1^{\psi_1},2^{\psi_1},4^{\psi_{234}},3^{\psi_{234}})\,\lambda^1_\beta\lambda^2_\alpha\tl^3_{\dot\alpha}\tl^4_{\dot\beta} \\ 
 -\, A(1^{\psi_1},3^{\psi_{234}},4^{\psi_{234}},2^{\psi_{1}})\,\lambda^1_\alpha\lambda^2_\beta \tl^3_{\dot\beta} \tl^4_{\dot\alpha} - A(1^{\psi_1},4^{\psi_{234}},3^{\psi_{234}},2^{\psi_{1}})\,\lambda^1_\alpha\lambda^2_\beta  \tl^3_{\dot\alpha} \tl^4_{\dot\beta}\bigg] \ , 
\end{split}
\end{align}
where the  relevant four-fermion amplitudes are 
\begin{align}
&A(1^{\psi_1},2^{\psi_1},3^{\psi_{234}},4^{\psi_{234}})\,
\,=\,-\,  \frac{\b{34}^2}{\b{23}\b{41}}\ , \nonumber \\
&A(1^{\psi_1},2^{\psi_1},4^{\psi_{234}},3^{\psi_{234}})\,
\,=\,- \, \frac{\b{34}^2}{\b{24}\b{31}}\, , \nonumber \\
&A(1^{\psi_1},3^{\psi_{234}},4^{\psi_{234}},2^{\psi_{1}}) \,=\, \frac{\b{34}^2}{\b{13}\b{42}}\, , \nonumber \\
&A(1^{\psi_1},4^{\psi_{234}},3^{\psi_{234}},2^{\psi_{1}})\,=\,  \frac{\b{34}^2}{\b{14}\b{32}}\ .
\label{ffa}
\end{align}
Using \eqref{ffa}, we can rewrite   \eqref{eq:contraction-fermions-2} as
\begin{align}
\frac{1}{4}\left[\frac{(\ell_2\bar{\ell_1})_{\alpha\beta}(\bar{\ell_4}\ell_3)_{\dot{\alpha} \dot{\beta}}+(\ell_1\bar{\ell_2})_{\alpha\beta}(\bar{\ell_3}\ell_4)_{\dot{\alpha} \dot{\beta}}}{2(\ell_2\cdot\ell_3)}\,+\, \ell_1\leftrightarrow \ell_2\right]\ .
\end{align}
The term with $\ell_1\leftrightarrow \ell_2$ is simply a relabelling of the integration variables, and we conclude  that the one-loop integrand  is given by
\begin{align}
{1\over 2} \left[\frac{(\ell_2\bar{\ell_1})_{\alpha\beta}(\bar{\ell_4}\ell_3)_{\dot{\alpha} \dot{\beta}}+(\ell_1\bar{\ell_2})_{\alpha\beta}(\bar{\ell_3}\ell_4)_{\dot{\alpha} \dot{\beta}}}{2(\ell_2\cdot\ell_3)}\right]\equiv 
N_{\alpha\beta\dot{\alpha}\dot{\beta}}\ .
\end{align}
Thus we have to consider the cut-integral 
\beq
\label{111}
I_{\alpha\beta\dot{\alpha}\dot{\beta}} := \int\!\! d^4\ell_1  d^4\ell_3 \,
\delta^{(+)}(\ell_1^2) \,  
\delta^{(+)}(\ell_3^2) \,  
\delta^{(+)} \left((L-\ell_1)^2\right) \,  
\delta^{(+)} \left((L+\ell_3)^2\right) \,  
\ \cdot \ N_{\alpha\beta\dot{\alpha}\dot{\beta}}\ . 
\eeq
It depends on only one scale $L$, hence it has the form 
\begin{align}
\label{eq:PV-2-fermions}
\,I_{\alpha\beta\dot{\alpha}\dot{\beta}}\,&=\,A\,L^2\epsilon_{\alpha\beta} \epsilon_{\dot{\alpha} \dot{\beta}}\,+\,B\,(L_{\alpha\dot{\alpha}}L_{\beta\dot{\beta}} + L_{\alpha\dot{\beta}}L_{\beta\dot{\alpha}})\ .
\end{align}
Contracting \eqref{111} and \eqref{eq:PV-2-fermions}
with $\epsilon^{\alpha\beta}\epsilon^{\dot{\alpha} \dot{\beta}}$ and $(\bar{L}^{\dot\alpha\alpha}\bar{L}^{\dot{\beta}\beta} + \bar{L}^{\dot{\beta}\alpha}\bar{L}^{\dot{\alpha}\beta})$ we can solve for the coefficients $A$ and $B$. 
The result for the corresponding UV-divergent parts is 
\beq
A_{\rm UV}\, =\, 0\ , \qquad  B_{\rm UV}\, =\, 1/6\ . 
\eeq
At this point we lift the four cut propagators to full propagators, so
that the cut double bubble becomes a full double-bubble integral. The
conclusion is then that the UV-divergent part of the integral
representing the two-fermion $\to$ two-fermion process is a
double bubble with coefficient
\begin{align}
\frac{1}{6}\,(L_{\alpha\dot{\alpha}}L_{\beta\dot{\beta}} + L_{\alpha\dot{\beta}}L_{\beta\dot{\alpha}})\ . 
\end{align}
This result has to be compared with the planar contractions at  tree level, 
\begin{align}
\label{eq:4-fermion-tree}
\begin{split}
I^{\rm tree}_{\alpha\beta\dot{\alpha}\dot{\beta}}\,&:=\,
\int\frac{d^DL_1}{(2\pi)^D}\frac{L_{1\,\alpha\dot{\beta}}(L-L_1)_{\beta\dot{\alpha}}}{L^2_1(L-L_1)^2}\ .
\end{split}
\end{align}
After a similar PV reduction of the $L_1$ integration in \eqref{eq:4-fermion-tree}, we find that $I^{\rm tree}_{\alpha\beta\dot{\alpha}\dot{\beta}}$  is given by a scalar (single) bubble with coefficient
\begin{align}
\frac{1}{6}\big(- L^2 \epsilon_{\alpha\beta} \epsilon_{\dot{\alpha}\dot{\beta}}\, + \, 
L_{\alpha\dot{\beta}}L_{\beta\dot{\alpha}}\big) \ , 
\end{align}
which using $L_{\alpha \dot\alpha} L_{\beta \dot\beta} - L_{\beta \dot\alpha} L_{\alpha \dot\beta} = L^2 \eps_{\alpha \beta} \eps_{\dot\alpha \dot\beta}$ 
can be rewritten as 
\begin{align}
{1\over 4} \left[  -  L^2 \, \epsilon_{\alpha\beta} \epsilon_{\dot{\alpha}\dot{\beta}}\, + \, 
{1\over 3} \big( L_{\alpha\dot{\alpha}}L_{\beta\dot{\beta}}+ L_{\beta\dot{\alpha}}L_{\alpha\dot{\beta}}\big)\right]  \ . 
\end{align}
This is the ``identity" or $\left\{\begin{smallmatrix}\alpha\,\beta\\ \cr \alpha\,\beta\end{smallmatrix}\right\}
$. 
The permutation is obtained by swapping $\dot\alpha$ and $\dot\beta$, or $\left\{\begin{smallmatrix}\alpha\,\beta\\ \cr \beta\,\alpha\end{smallmatrix}\right\}
$.  Thus, we can write: 
\beqa
\left\{\begin{smallmatrix}\alpha\,\beta\\ \cr \alpha\,\beta\end{smallmatrix}\right\}
: \quad &&{1\over 4} \left[\, -L^2 \, \epsilon_{\alpha\beta} \epsilon_{\dot{\alpha}\dot{\beta}}\, + \, 
{1\over 3} \big( L_{\alpha\dot{\alpha}}L_{\beta\dot{\beta}}+ L_{\beta\dot{\alpha}}L_{\alpha\dot{\beta}}\big)\right]\, , 
\\ 
\left\{\begin{smallmatrix}\alpha\,\beta\\ \cr \beta\,\alpha\end{smallmatrix}\right\}
: \quad &&{1\over 4} \left[   \, L^2 \, \epsilon_{\alpha\beta} \epsilon_{\dot{\alpha}\dot{\beta}}\, + \, 
{1\over 3} \big( L_{\alpha\dot{\alpha}}L_{\beta\dot{\beta}}+ L_{\beta\dot{\alpha}}L_{\alpha\dot{\beta}}\big)\right]\, .
\eeqa
In this language, the tree-level contraction is represented as  
\beq
\left\{\begin{smallmatrix}\alpha\,\beta\\ \cr \alpha\,\beta\end{smallmatrix}\right\}
\ .
\eeq
Hence, also reinstating powers of the 't Hooft coupling,  we obtain  that the term in the spin-chain Hamiltonian corresponding to the two-fermion $\to$ two-fermion process is
\beq
\label{2f}
{\lambda \over 8 \pi^2} \left( \left\{\begin{smallmatrix}\alpha\,\beta\\ \cr \alpha\,\beta\end{smallmatrix}\right\}
 \ + \ \left\{\begin{smallmatrix}\alpha\,\beta\\ \cr \beta\,\alpha\end{smallmatrix}\right\} \right)
  \ , 
\eeq
in agreement with the corresponding terms in \eqref{dilsu23}. In conclusion, putting together the purely scalar result of Section 2, \eqref{pss}, as well as the results  \eqref{1f1s} and \eqref{2f} for the two-fermion two-scalar and four-fermion cases, we have confirmed the complete expression \eqref{dilsu23} for the spin-chain Hamiltonian in the $SU(2|3)$ sector.

\section{Conclusions} 

We would like to summarise some of the key points of our paper, 
compare with other recent on-shell approaches
and finally make a few suggestions for  future research. 

The calculation of the dilatation operator in $\mathcal{N}=4$ SYM has
been revisited in recent months using on-shell and twistor (string) inspired approaches. 
In \cite{Wilhelm:2014qua} the complete one-loop dilatation operator was obtained by calculating form factors for generic single-trace operators using generalised unitarity, making interesting contact with earlier work of  
\cite{Zwiebel:2011bx}. In particular, the integral form for the dilatation operator in \cite{Zwiebel:2011bx} is mapped to a phase-space integral, which appears naturally in a unitarity-based approach. 
The calculation of two-loop form factors using unitarity was also employed to obtain  the two-loop anomalous dimension of the Konishi operator in  \cite{Nandan:2014oga}.

On the other hand, in \cite{Koster:2014fva},  twistor-space  MHV diagrams were used to find the dilatation operator in the $SO(6)$ sector at one loop directly from two-point correlators, leading to the 
position-space form of the correlator as found by \cite{Minahan:2002ve}.
In a closely related approach, in \cite{Brandhuber:2014pta} MHV diagrams in momentum space were shown to 
reproduce the $SO(6)$ one-loop dilatation operator. In momentum space  the calculation gives a single-scale two-loop integral, which after Fourier transform gives the expected result. Perhaps one interesting difference between these two MHV-based approaches is that the twistor-space computation requires an additional line-splitting regularisation of the operator. Finally, in the current paper we have simplified the calculation of \cite{Brandhuber:2014pta} considerably by directly applying   generalised unitarity to the calculation of the two-point functions, which should have obvious  generalisations to higher loops. 

In comparing the two main lines of approach, using form factors or the two-point correlators, one notices the following main points. In order to extract $L$-loop anomalous dimensions from form factors, an $L$-loop calculation is required, while for the two-point correlators in momentum space in principle $2 L$-loop integrals can appear. However, form factors also have (universal) infrared divergences which need to be disentangled from the UV divergences, and with increasing loop order one obtains integrals with an increasing number of scales. 
In the case of two-point correlators, one has the advantage of only having to consider single-scale integrals, albeit at higher-loop order in momentum space, and one never encounters infrared divergences. More work is clearly needed to determine which method is more efficient, but we think that all approaches have their own merits and will shed interesting new light on the problem of calculating the dilatation operator, and hopefully lead to a proof of the integrability conjecture.

Let us now make some concluding comments on our findings and point out future directions.

{\bf 1. } We stress that in our method no integrals are computed  at any stage. We only identify coefficients of a single quadruple-cut integral, and from its UV divergence we read off the dilatation operator. Some PV reductions are performed on shell, which are also of algebraic nature. 

{\bf 2.} At one loop, quadruple cuts are sufficient to determine the dilatation operator. This is related to the fact that at this loop order only four fields are connected (and thus the four propagators we cut are always present).  This circumstance is not related to the presence of maximal supersymmetry, and hence we can envisage obvious applications to  theories with  $\cN<4$ or even  no supersymmetry. 

{\bf 3.} In the calculations presented here (as well as in \cite{Brandhuber:2014pta}) we have made use of amplitudes with scalars and fermions. The use of gluon amplitudes remains as a future direction of research, and we expect these to be  relevant for the study of the $SL(2)$ sector as well as for single-trace operators made of field strengths in QCD \cite{Ferretti:2004ba}.  

{\bf 4.} Clearly the application of our method to higher loops is a crucial testing ground -- the ultimate goal being proving  integrability at higher loops (rather than assuming it).

{\bf 5.} It would also be interesting to perform the $SU(2|3)$ calculation with MHV diagrams, thus extending the approach of \cite{Brandhuber:2014pta} to fermions. 

We will come back to these issues in future work. 

\section*{Acknowledgements}

It is a pleasure  to thank Lorenzo Bianchi, Valentina Forini, Jan Plefka  and Matthias Staudacher for  interesting discussions. GT thanks the Institute for Physics and  IRIS Adlershof at Humboldt University, Berlin, for their warm hospitality and support.
This work was supported by the Science and Technology Facilities Council Consolidated Grant ST/L000415/1  
{\it ``String theory, gauge theory \& duality". }

\bibliographystyle{utphys}
\bibliography{dilatation}
\end{document}